\documentstyle[12pt]{article}

\def\df #1. #2\par{\medbreak
  \noindent{{\tt {\bf Definition #1.}}\enspace}{\sl#2\par}%
  \ifdim\lastskip<\medskipamount \removelastskip\penalty55\medskip\fi}

\def\theorem #1. #2\par{\medbreak
  \noindent{\tt {\bf Theorem #1.}\enspace}{\sl#2\par}%
  \ifdim\lastskip<\medskipamount \removelastskip\penalty55\medskip\fi}

\def\lemma #1. #2\par{\medbreak
  \noindent{\tt {\bf Lemma #1.}\enspace}{\sl#2\par}%
  \ifdim\lastskip<\medskipamount \removelastskip\penalty55\medskip\fi}

\def\proof{\medbreak\noindent{\bf Proof}}

\def\dsl{\raise.15ex\hbox{/}\kern-.65em\nabla}

\def\pi{\noindent{\bf Physical interpretation}. }

\def\la{\Lambda}

\def\D{{\cal D}}

\def\M{{\cal M}}

\def\L{{\cal L}}
\def\V{{\cal V}}

\def\t{\tilde}
\def\be{\begin{equation}}
\def\ee{\end{equation}}

\def\tr{{\rm Tr}}
\def\ww{\wedge\ldots\wedge}

\def\det{{\rm det}}
\def\even{{\rm even}}
\def\odd{{\rm odd}}

\def\kfac{\frac{1}{k!}}

\def\Spin{{\rm Spin}_\V}
\def\com{{\rm Com}}
\def\exp{{\rm exp}}
\def\diag{{\rm diag}}
\def\s{\stackrel}

\begin{document}

\title{The tensor Dirac equation in Riemannian space}

\author{N.G.Marchuk \thanks{Research supported by the Russian Foundation
for Basic Research, grant 00-01-00224, and by the Royal Society.}}


\maketitle

Steklov Mathematical Institute, Gubkina st.8,
Moscow 117966, Russia;
nmarchuk@mi.ras.ru;
http://www.orc.ru/\~{}nmarchuk
\vskip 1cm


\begin{abstract}
We suggest a tensor equation on Riemannian manifolds which can be
considered as a generalization of the Dirac equation for the electron.
The tetrad formalism is not used. Also we suggest a new form of the
tensor Dirac equation with a Spin(1,3) gauge symmetry in Minkowski
space.
\end{abstract}


\tableofcontents

\vskip 1cm

In this paper, following \cite{Marchuk1},\cite{Marchuk2}, we consider
the tensor Dirac equation (\ref{main}) on Riemannian
manifolds. Our approach leads to  a new form of the
tensor Dirac equation with a Spin(1,3) gauge symmetry in Minkowski
space.

The research was carried out while the author was visiting at Bath
University. The author is grateful to Professor D.Vassiliev, Dr. A.King,
and Dr. F.Burstall for useful discussions and for hospitality.

\bigskip
\section{ Differential forms on Riemannian manifolds.}
Let $\M$ be a four dimensional differentiable manifolds covered by a
system of coordinates $x^\mu$. Greek indices run over (0,1,2,3).
Summation convention over repeating indices is assumed. We consider
atlases on $\M$ consisted of one chart. Suppose that there is a smooth
twice covariant tensor field (a metric tensor) with components
$g_{\mu\nu}=g_{\mu\nu}(x)$, $x\in\M$ such that
\begin{itemize}
\item $g_{\mu\nu}=g_{\nu\mu}$;
\item $g = \det\|g_{\mu\nu}\|<0$ for all $x\in\M$;
\item The signature of the matrix $\|g_{\mu\nu}\|$ is equal to $-2$.
\end{itemize}

The matrix $\|g^{\mu\nu}\|$ composed from contravariant components of
the metric tensor is the inverse matrix to $\|g_{\mu\nu}\|$.
The full set of $\{\M,g_{\mu\nu}\}$ is called {\sl an elementary
Riemannian manifolds} (with one chart atlases) and is denoted by $\V$.

Let $\Lambda^k$ be the sets of exterior differential
forms of rank $k=0,1,2,3,4$ on $\V$ (covariant antisymmetric tensor fields) and
$$
\la=\la^0\oplus\ldots\oplus\la^4=\la^\even\oplus\la^\odd,
$$
$$
\la^\even=\la^0\oplus\la^2\oplus\la^4,\quad
\la^\odd=\la^1\oplus\la^3.
$$
Elements of $\la$ are called (nonhomogeneous) {\sl differential forms}
and elements of $\la^k$ are called {\sl $k$-forms} or differential forms
of rank $k$. The set of smooth scalar functions on $\V$ (invariants) is
identified with the set of $0$-forms $\la^0$. A $k$-form $U\in\la^k$ can
be written as
\begin{equation}
U=\kfac u_{\nu_1\ldots \nu_k}dx^{\nu_1}\ww dx^{\nu_k}=
\sum_{\mu_1<\cdots<\mu_k} u_{\mu_1\ldots \mu_k}dx^{\mu_1}\ww dx^{\mu_k},
\label{k-form}
\end{equation}
where $u_{\nu_1\ldots \nu_k}=u_{\nu_1\ldots \nu_k}(x)$ are real valued
components of a covariant antisymmetric
($u_{\nu_1\ldots \nu_k}=u_{[\nu_1\ldots \nu_k]}$) tensor field.
Differential forms from $\la$ can be written as linear combinations of
the 16 basis differential forms
\begin{equation}
1,dx^\mu,dx^{\mu_1}\wedge dx^{\mu_2},\ldots,dx^{0}\wedge\ldots\wedge
dx^3,
\quad
\mu_1<\mu_2<\ldots.
\label{basis}
\end{equation}
The exterior multiplication of differential forms is defined in the
usual way.
If $U\in\la^r,V\in\la^s$, then
$$
U\wedge V=(-1)^{rs}V\wedge U\in\la^{r+s}.
$$
In this paper we consider changes of coordinates with positive Jacobian
and do not distinguish tensors and pseudotensors.

Consider the Hodge star operator
$\star\,:\,\Lambda^k\to\Lambda^{4-k}$. If $U\in\Lambda^k$
has the form (\ref{k-form}), then
$$
\star U=
\frac{1}{k!(4-k)!}\sqrt{-g}\,\varepsilon_{\mu_1\ldots\mu_4}u^{\mu_1\ldots\mu_k}
dx^{\mu_{k+1}}\ww dx^{\mu_4},
$$
where
$u^{\mu_1\ldots\mu_k}=g^{\mu_1\nu_1}\ldots g^{\mu_k\nu_k}u_{\nu_1\ldots\nu_k}
$,
$\varepsilon_{\mu_1\ldots\mu_4}$ is the sign of the permutation
$(\mu_1\ldots\mu_4)$, and $\varepsilon_{0123}=1$. It is easy to prove
that for $U\in\la^k$
$$
\star(\star U)=(-1)^{k+1}U.
$$

Further on we consider the bilinear operator
$\com\,:\,\Lambda^2\times\Lambda^2\to\Lambda^2$ such that
\begin{eqnarray*}
&&\com(\frac{1}{2}a_{\mu_1\mu_2}dx^{\mu_1}\wedge dx^{\mu_2},
\frac{1}{2}b_{\nu_1\nu_2}dx^{\nu_1}\wedge dx^{\nu_2})=
\frac{1}{2}a_{\mu_1\mu_2}b_{\nu_1\nu_2}(-g^{\mu_1 \nu_1}dx^{\mu_2}\wedge dx^{\nu_2}\\
&&-g^{\mu_2 \nu_2}dx^{\mu_1}\wedge dx^{\nu_1}
+g^{\mu_1 \nu_2}dx^{\mu_2}\wedge dx^{\nu_1}+g^{\mu_2
\nu_1}dx^{\mu_1}\wedge dx^{\nu_2})
\end{eqnarray*}
Evidently, $\com(U,V)=-\com(V,U)$.

Now we define the Clifford multiplication of differential forms with the aid of
the following formulas (see formulas for the space dimensions 2 and 3 in
\cite{Marchuk1}):
\begin{eqnarray*}
\s{0}{U}\s{k}{V}&=&\s{k}{V}\s{0}{U}=\s{0}{U}\wedge\s{k}{V}=\s{k}{V}\wedge\s{0}{U},\\
\s{1}{U}\s{k}{V}&=&\s{1}{U} \wedge  \s{k}{V}-\star (\s{1}{U} \wedge  \star \s{k}{V}),\\
\s{k}{U}\s{1}{V}&=&\s{k}{U} \wedge  \s{1}{V}+\star (\s{k}{U} \wedge  \star \s{1}{V}),\\
\s{2}{U}\s{2}{V}&=&\s{2}{U}\wedge\s{2}{V}+
\star(\s{2}{U}\wedge\star\s{2}{V})+\frac{1}{2}\com(\s{2}{U},\s{2}{V}),\\
\s{2}{U}\s{3}{V}&=&\star \s{2}{U} \wedge  \star \s{3}{V}-\star (\s{2}{U} \wedge  \star \s{3}{V}),\\
\s{2}{U}\s{4}{V}&=&\star \s{2}{U} \wedge  \star \s{4}{V},\\
\s{3}{U}\s{2}{V}&=&-\star \s{3}{U} \wedge  \star \s{2}{V}-\star (\star \s{3}{U} \wedge  \s{2}{V}),\\
\s{3}{U}\s{3}{V}&=&\star \s{3}{U} \wedge  \star \s{3}{V}+\star (\s{3}{U} \wedge  \star \s{3}{V}),\\
\s{3}{U}\s{4}{V}&=&\star \s{3}{U} \wedge  \star \s{4}{V},\\
\s{4}{U}\s{2}{V}&=&\star \s{4}{U} \wedge  \star \s{2}{V},\\
\s{4}{U}\s{3}{V}&=&-\star \s{4}{U} \wedge  \star \s{3}{V},\\
\s{4}{U}\s{4}{V}&=&-\star \s{4}{U} \wedge  \star \s{4}{V},
\end{eqnarray*}
where ranks of differential forms are denoted as $\s{k}{U}\in\Lambda^k$ and
$k=0,1,2,3,4$.
From this definition we may obtain some properties of the Clifford
multiplication of differential forms.
\begin{description}
\item[1.] If $U,V\in\Lambda$, then $UV\in\Lambda$.
\item[2.] The axioms of associativity and distributivity are satisfied
for the Clifford multiplication.
\item[3.] $dx^\mu dx^\nu=dx^\mu\wedge dx^\nu+g^{\mu\nu},\quad
dx^\mu dx^\nu+dx^\nu dx^\mu=2 g^{\mu\nu}$.
\item[4.] If $U,V\in\Lambda^2$, then $\com(U,V)=UV-VU$.
\end{description}

Let us define the trace of a differential form as a linear operation
$\tr\,:\,\la\to\la^0$ such that
$$
\tr(1)=1,\quad
\tr(dx^{\mu_1}\ww dx^{\mu_k})=0\quad\hbox{for}\quad k=1,2,3,4.
$$
The reader can easily prove
that
$$
\tr(UV-VU)=0,\quad \tr(V^{-1}UV)=\tr\,U,\quad U,V\in\la.
$$

Let us define an involution $*\,:\,\la^k\to\la^k$. By definition,
put
$$
U^*=(-1)^{\frac{k(k-1)}{2}}U,\quad U\in\la^k.
$$
It is readily seen that
$$
U^{**}=U,\quad (UV)^*=V^* U^*,\quad U,V\in\la.
$$

Now we can define the spinor group
$$
\Spin = \{S\in \la^\even\,:\,S^*S=1\}.
$$

\bigskip

\section{ Tensors with values in $\la^k$.}
Let
$$
u^{\lambda_1\ldots \lambda_r}_{\mu_1\ldots \mu_k \nu_1\ldots \nu_s}(x)
=u^{\lambda_1\ldots \lambda_r}_{[\mu_1\ldots \mu_k]\nu_1\ldots \nu_s}(x),\quad x\in\V
$$
be components of a tensor field of rank
$(r,k+s)$ antisymmetric with respect to the first
$k$ covariant indices. One may consider the following objects:
\begin{equation}
U^{\lambda_1\ldots \lambda_r}_{\nu_1\ldots \nu_s}=
\kfac u^{\lambda_1\ldots \lambda_r}_{\mu_1\ldots \mu_k \nu_1\ldots
\nu_s}\,dx^{\mu_1}\wedge\ldots\wedge dx^{\mu_k}
\label{index:form}
\end{equation}
which are formally written as
$k$-forms. Under a change of coordinates
$(x)\to(\tilde x)$ the values
(\ref{index:form}) transform as components of a tensor field of
rank $(r,s)$
\begin{equation}
\tilde U^{\alpha_1\ldots \alpha_r}_{\beta_1\ldots \beta_s}=
q^{\nu_1}_{\beta_1}\ldots q^{\nu_s}_{\beta_s} p^{\alpha_1}_{\lambda_1}\ldots p^{\alpha_r}_{\lambda_r}
U^{\lambda_1\ldots \lambda_r}_{\nu_1\ldots
\nu_s},\quad q^\nu_\beta=\frac{\partial x^\nu}{\partial\tilde x^\beta},
\quad p^\alpha_\lambda=\frac{\partial\tilde x^\alpha}{\partial x^\lambda}.
\label{index:form1}
\end{equation}
The objects (\ref{index:form}) are called tensors of rank $(r,s)$ with
values in $\la^k$. We write this as
$$
U^{\lambda_1\ldots \lambda_r}_{\nu_1\ldots \nu_s}\in\la^k\top^r_s.
$$
Elements of $ \la^0\top^r_s$ are ordinary tensors of rank $(r,s)$ on
$\V$.
For $U_\mu\in\la^k\top_1$ we have
$$
dx^\mu U_\mu\in\la^{k+1}\oplus\la^{k-1}.
$$
\bigskip

\section{ The covariant derivatives $\nabla_\mu$.}
On Riemannian manifolds $\V$ the Christoffel symbols
${\Gamma^\lambda}_{\mu\nu}={\Gamma^\lambda}_{\nu\mu}$ (Levi-Chivita
connectedness components) are defined with the aid of the metric tensor
\begin{equation}
{\Gamma^\lambda}_{\mu\nu}=
\frac{1}{2}g^{\lambda\kappa}(\partial_\mu g_{\mu\kappa}+
\partial_\nu g_{\mu\kappa}-\partial_\kappa g_{\mu\nu}).
\end{equation}
Let us remind the definition
of covariant derivatives $\nabla_\mu$ acting on tensor fields on $\V$ by
the following rules ($\partial_\mu=\partial/\partial x^\mu$):
\medskip

\noindent 1. If $t=t(x),\ x\in\V$ is a scalar function (invariant), then
$$
\nabla_\mu t=\partial_\mu t.
$$

\noindent 2. If $t^\nu$ is a vector field on $\V$, then
$$
\nabla_\mu t^\nu\equiv t^\nu_{;\mu}=\partial_\mu t^\nu +
{\Gamma^\nu}_{\mu\lambda} t^\lambda.
$$

\noindent 3. If $t_\nu$ is a covector field on $\V$, then
$$
\nabla_\mu t_\nu\equiv t_{\nu;\mu}=\partial_\mu t_\nu - {\Gamma^\lambda}_{\mu\nu} t_\lambda.
$$

\noindent 4. If $u=u^{\nu_1\ldots \nu_k}_{\lambda_1\ldots \lambda_l},\
v=v^{\nu_1\ldots \nu_r}_{\lambda_1\ldots \lambda_s}$ are tensor fields on $\V$, then
$$
\nabla_\mu(u\otimes v)=(\nabla_\mu u)\otimes v + u\otimes\nabla_\mu v.
$$
\medskip

With the aid of these rules it is easy to calculate covariant
derivatives of arbitrary tensor fields.
Also, it is easy to check the correctness of the following formulas:
$$
\nabla_\mu g_{\nu\lambda}=0,\quad \nabla_\mu g^{\nu\lambda}=0,
\quad \nabla_\mu\delta^\nu_\lambda=0.
$$

\bigskip

\section{ The Clifford derivatives $\Upsilon_\mu$.}
Let us define the Clifford derivatives $\Upsilon_\mu$
(Upsilon), which act on tensors from $\la\top^r_s$ by the following rules:
\medskip

\noindent 1. If $t_{\nu_1\ldots \nu_s}^{\epsilon_1\ldots \epsilon_r}$
is a covariant tensor field on
$\V$ of rank
$(r,s)$, then
$$
\Upsilon_\mu t_{\nu_1\ldots \nu_s}^{\epsilon_1\ldots \epsilon_r}=
\partial_\mu t_{\nu_1\ldots \nu_s}^{\epsilon_1\ldots \epsilon_r}.
$$

\noindent 2. $\Upsilon_\mu dx^\nu = -{\Gamma^\nu}_{\mu\lambda} dx^\lambda$.
\smallskip

\noindent 3. If $U,V\in\la$ and $UV$ is the Clifford product of
differential forms, then
$$
\Upsilon_\mu(UV)=(\Upsilon_\mu U)V+ U\Upsilon_\mu V.
$$
\medskip

With the aid of these rules it is easy to calculate how operators
$\Upsilon_\mu$ act on arbitrary tensor from
$\la\top_s^r$.

If $U\in\la^k$, written as (\ref{k-form}), then
\begin{equation}
\Upsilon_\mu U=\kfac u_{\nu_1\ldots \nu_k;\mu}dx^{\nu_1}\wedge\ldots\wedge dx^{\nu_k}.
\label{UpsilonU}
\end{equation}
That means $\Upsilon_\mu\,:\,\la^k\to\la^k\top_1$.
The formula (\ref{UpsilonU}) indicate the connection between operators
$\Upsilon_\mu$ and $\nabla_\mu$.

If $U_{\nu_1\ldots\nu_s}^{\epsilon_1\ldots \epsilon_r}\in\la\top_s^r$
and $r+s>0$, then the values
$\Upsilon_\mu U_{\nu_1\ldots\nu_s}^{\epsilon_1\ldots \epsilon_r}$
are not the
components of a tensor (when the curvature is nonzero).
In what follows we do not use the Clifford derivatives $\Upsilon_\mu$ as
isolated operators acting on tensors from $\la\top_s^r,r+s>0$. But we use
them as building blocks of operators, that map tensors to tensors. For
example, if $B_\mu\in\la^2\top_1$, then the expression
$$
\Upsilon_\mu B_\nu-\Upsilon_\nu B_\mu-[B_\mu,B_\nu]
$$
is a tensor from $\la^2\top_2$.

Consider the change of coordinates $(x)\to(\t x)$
$$
p^\mu_\nu=\frac{\partial\t x^\mu}{\partial x^\nu},\quad
q^\mu_\nu=\frac{\partial x^\mu}{\partial\t x^\nu},\quad
dx^\mu=q^\mu_\nu d\t x^\nu,
$$
where $p^\mu_\nu,q^\mu_\nu$ are functions of  $x\in\V$. Then the
Clifford derivatives
$\Upsilon_\nu$ in coordinates $x^\mu$ related to
the Clifford derivatives $\t\Upsilon_\nu$ in coordinates
$\t x^\mu$ by the formula
\begin{equation}
\Upsilon_\nu=p^\mu_\nu \t\Upsilon_\mu
\label{tilde:nabla}
\end{equation}
exact the same as formula for partial derivatives
$\partial_\nu=p^\mu_\nu \t\partial_\mu$,
where $\partial_\nu=\frac{\partial}{\partial
x^\nu}$, $\t\partial_\nu=\frac{\partial}{\partial\t x^\nu}$.
The proof of this formula is followed from
the transformation rule of Christoffel symbols.

The main properties of the operators $\Upsilon_\mu$ are listed
below.
\medskip

1) $\Upsilon_\mu(U^*)=(\Upsilon_\mu U)^*$ for $U\in\la$.

2) $\Upsilon_\mu(\star U)=\star(\Upsilon_\mu U)$ for $U\in\la$.

3) $\Upsilon_\mu(\tr\,U)=\tr(\Upsilon_\mu U)$ for $U\in\la$.

\medskip

From the formula $\Upsilon_\mu dx^\lambda=-{\Gamma^\lambda}_{\mu\nu}dx^\nu$ we get
\begin{equation}
(\Upsilon_\mu\Upsilon_\nu-\Upsilon_\nu\Upsilon_\mu)dx^\lambda=
-{R^\lambda}_{\rho\mu\nu}dx^\rho,
\label{R1}
\end{equation}
where
\begin{equation}
{R^\kappa}_{\lambda\mu\nu}=
\partial_\mu {\Gamma^\kappa}_{\nu\lambda}-\partial_\nu
{\Gamma^\kappa}_{\mu\lambda}+
{\Gamma^\kappa}_{\mu\eta}{\Gamma^\eta}_{\nu\lambda}-
{\Gamma^\kappa}_{\nu\eta}{\Gamma^\eta}_{\mu\lambda}
\label{R2}
\end{equation}
is the rank (1,3) tensor, known as the curvature tensor (or Riemannian
tensor).
Consider the antisymmetric tensor from $\Lambda^2\top_2$ such that
$$
C_{\mu\nu}=\frac{1}{2}R_{\alpha\beta\mu\nu}dx^\alpha\wedge dx^\beta.
$$

\theorem 1. For all $U\in\la$
\begin{equation}
(\Upsilon_\mu \Upsilon_\nu-\Upsilon_\nu \Upsilon_\mu)U=
\frac{1}{2}[C_{\mu\nu},U].
\label{Gen}
\end{equation}
The proof is by direct calculation.
\bigskip

If $U$ is invertible (w.r.t. Clifford multiplication) differential form
from $\la$,
then
the relation (\ref{Gen}) can be written as
\begin{equation}
\frac{1}{2}C_{\mu\nu}=U^{-1}(\frac{1}{2}C_{\mu\nu})U-
U^{-1}(\Upsilon_\mu \Upsilon_\nu-\Upsilon_\nu \Upsilon_\mu)U.
\label{Gen1}
\end{equation}

Let $B_\mu\in\Lambda^2\top_1$ be a tensor such that
\begin{equation}
\Upsilon_\mu B_\nu-\Upsilon_\nu B_\mu-[B_\mu,B_\nu]=
\frac{1}{2}C_{\mu\nu}.
\label{BG}
\end{equation}
\bigskip

The existence of solutions of this equation must be investigated.
\bigskip

\theorem 2. If $B_\mu\in\Lambda^2\top_1$ satisfy (\ref{BG}) and
$S\in \Spin$, then $B_\mu^\prime=S^{-1}B_\mu S-S^{-1}\Upsilon_\mu S$
also satisfy (\ref{BG}).

\proof. It is easily shown that the formula (\ref{BG}) is invariant
under the following gauge transformation with $\Spin$ symmetry group:
\begin{eqnarray}
B_\mu &\to& B_\mu^\prime=S^{-1}B_\mu S-S^{-1}\Upsilon_\mu S,
\label{gauge:trans}\\
\frac{1}{2}C_{\mu\nu} &\to& \frac{1}{2}C_{\mu\nu}^\prime=
S^{-1}(\frac{1}{2}C_{\mu\nu})S-
S^{-1}(\Upsilon_\mu \Upsilon_\nu-\Upsilon_\nu \Upsilon_\mu)S,
\nonumber
\end{eqnarray}
where $S\in \Spin$. Substituting $U=S$ to the formula (\ref{Gen1}), we
obtain $C_{\mu\nu}^\prime\equiv C_{\mu\nu}$. This completes the proof.
\bigskip

Now we may connect our construction to the following linear system of
differential equations:
\begin{equation}
\Upsilon_\mu U-[B_\mu,U]=0,
\label{eq}
\end{equation}
where $U\in\Lambda$ is an unknown differential form and $B_\mu$
satisfies (\ref{BG}).
 There are three things to be said
about this system of equations.

The first. The equations (\ref{eq}) are invariant under the gauge
transformation
$$
U\to S^{-1}US,\quad B_\mu\to S^{-1}B_\mu S-S^{-1}\Upsilon_\mu S,
\quad S\in \Spin.
$$

The second. From  (\ref{eq}) we
may get the following equalities:
\begin{equation}
\Upsilon_\mu(\Upsilon_\nu U-[B_\nu,U])-
\Upsilon_\nu(\Upsilon_\mu U-[B_\mu,U])=0.
\label{eq1}
\end{equation}
If $U\in\la$ satisfies (\ref{eq}), then $U$ also satisfies (\ref{eq1}).
Hence equalities (\ref{eq1}) can be considered as  necessary conditions
for the existence of a solution of the equations (\ref{eq}). We see that
equalities (\ref{eq1}) are equivalent to the equalities
(\ref{Gen}), which are valid
for any $U\in\la$.

The third. If $U_1$ and $U_2$ satisfy (\ref{eq}), then $U_1 U_2$
and $U_1+U_2$ also
satisfy (\ref{eq}).

\bigskip

\section{ Equations for the electron on Riemannian manifolds.}
In \cite{JMP} we prove that in Minkowski space the Dirac equation for
the electron can be written as a tensor equation (the tensor Dirac
equation). Now we present the following system of tensor equations on Riemannian
manifolds, which can be considered as a generalization of the tensor Dirac
equation:
\begin{eqnarray}
&&dx^\mu(\Upsilon_\mu\Psi+\Psi I a_\mu+\Psi B_\mu)+m\Psi
HI=0,\nonumber\\
&&\Upsilon_\mu H=[B_\mu,H],\quad \Upsilon_\mu I=[B_\mu,I],\label{main}\\
&&H^2=1,\quad I^2=-1,\quad [H,I]=0,\nonumber
\end{eqnarray}
where
\begin{equation}
\Psi\in\la^\even,\quad I\in\la^2,\quad H\in\la^1,\quad a_\mu\in\la^0\top_1,
\quad B_\mu\in\la^2\top_1,
\label{main:cond}
\end{equation}
$m\geq0$ is a real constant, and $B_\mu$ satisfies
(\ref{BG}). We suppose that in (\ref{main})
the differential forms $\Psi,H,I$ are unknown and the tensors
$a_\mu,B_\mu$ are known.
\bigskip

\theorem 3. The system of equations (\ref{main}) is invariant under the
gauge transformation
\begin{equation}
\Psi\to\Psi\exp(\lambda I),\quad a_\mu\to a_\mu-\partial_\mu\lambda,
\quad (H,I,B_\mu)\to(H,I,B_\mu),
\label{trans1}
\end{equation}
where $\lambda=\lambda(x)\in\la^0$ and $\exp(\lambda I)=\cos\lambda+
I\sin\lambda$.
\par

\proof. Denote $S=\exp(\lambda I)$. We have
\begin{eqnarray}
\Upsilon_\mu S&=&(\Upsilon_\mu I)\sin\lambda+
I\partial_\mu\lambda(\cos\lambda+I\sin\lambda) \label{Ups:S}\\
&=&[B_\mu,I]\sin\lambda+SI\partial_\mu\lambda\nonumber
\end{eqnarray}
Multiplying the first equation in (\ref{main}) from the right by $S$
and denoting $\Psi^\prime=\Psi S$, we obtain
$$
dx^\mu(\Upsilon_\mu\Psi^\prime+\Psi^\prime
(-S^{-1}\Upsilon_\mu S+a_\mu I+S^{-1}B_\mu S))+m\Psi^\prime IH=0.
$$
If we substitute  $\Upsilon_\mu S$ from (\ref{Ups:S})
to this equation,
then we get
$$
-S^{-1}\Upsilon_\mu S+a_\mu I+S^{-1}B_\mu S=
(a_\mu-\partial_\mu\lambda)I+B_\mu.
$$
This completes the proof.
\bigskip

\theorem 4. The system of equations (\ref{main}) is invariant under the
gauge transformation
\begin{equation}
\Psi\to\Psi S,\quad H\to S^{-1}HS,\quad I\to S^{-1}IS,\quad
B_\mu\to S^{-1}B_\mu S-S^{-1}\Upsilon_\mu S,\quad a_\mu\to a_\mu,
\label{trans2}
\end{equation}
where $S\in \Spin$.\par

The proof is evident.

\bigskip

\section{ The conservative law and the Lagrangian.}
With the aid of the 1-form $H$ we define the operation of
conjugation
$$
\bar{U}=HU^*,\quad U\in\la.
$$

\lemma. Suppose $\Psi,H,I,a_\mu,B_\mu$ are chosen as in (\ref{main:cond})
and
$$
L=\Psi^*(dx^\mu(\Upsilon_\mu\Psi+\Psi I a_\mu+\Psi B_\mu)+m\Psi H I).
$$
Then  the conjugated differential form $\bar{L}$ can be written as
$$
\bar{L}=((\Upsilon_\mu\bar{\Psi}-a_\mu I\bar{\Psi}-B_\mu\bar{\Psi})dx^\mu-
mI H\bar{\Psi})\Psi.
$$
\par

Proof is by direct calculation.
\par

\theorem 5. Let $\Psi,H,I,a_\mu,B_\mu$ satisfy
(\ref{main}),(\ref{main:cond}) and
$j^\mu=\tr(\bar{\Psi}dx^\mu\Psi)$. Then
\begin{equation}
\partial_\mu(\sqrt{-g}\,j^\mu)=0.
\label{conserv}
\end{equation}
\par

The identity (\ref{conserv}) is called {\sl a conservative law} for the
equation (\ref{main}). The vector $j^\mu$ is called  {\sl a current}.

Proof. It can be checked that
$$
\tr(H(L+L^*))=\frac{\partial_\mu(\sqrt{-g}\,j^\mu)}{\sqrt{-g}}.
$$
For a solution of the equation (\ref{main}) we have $L=0$ and so we
obtain the conservative law (\ref{conserv}). This completes the proof.
\medskip

Let us define the Lagrangian (the Lagrangian density)
from which the main equation (\ref{main})
can be derived
\begin{eqnarray*}
\L&=&\tr(\sqrt{-g}HLI)\\
&=&\tr(\sqrt{-g}\bar{\Psi}(dx^\mu(\Upsilon_\mu\Psi+\Psi I a_\mu+\Psi
B_\mu)I - m\Psi H)
\label{lagr}
\end{eqnarray*}
Note that this Lagrangian is invariant under the gauge transformations
(\ref{trans1}) and (\ref{trans2}).

Using the variational principle
\cite{Bogolubov} we suppose that in the Lagrangian $\L$
the differential forms $\Psi$ and
$\bar{\Psi}$ are independent
and as variational variables we take 8
functions which are the coefficients of the differential form
$\bar{\Psi}$. The Lagrange-Euler equations with respect to these
variables give us the system of equations, which can be written in the
form (\ref{main}).
\bigskip


\section{ The operators $d,\delta,\Upsilon$.}
With the aid of Clifford derivatives one can define
three differential operators of first order, which map
$\la$ into $\la$. By definition, put
\begin{eqnarray*}
dV&=&dx^\mu\wedge\Upsilon_\mu V,\\
\Upsilon V&=&dx^\mu \Upsilon_\mu V,\\
\delta V&=&dV-\Upsilon V,
\end{eqnarray*}
for $V\in\la$.
Some of the properties of these operators are listed below.
\begin{itemize}
\item $d\,:\,\la^k\to\la^{k+1}$;
\item $d^2=0$;
\item $d(U\wedge V)=dU\wedge V+(-1)^k U\wedge dV$, $U\in\la^k, V\in\la$;
\item $\delta\,:\,\la^k\to\la^{k-1}$;
\item $\delta^2=0$;
\item $\delta U=\star d\star U$, $U\in\la$;
\item $\Upsilon\,:\,\la^k\to\la^{k-1}\oplus\la^{k+1}$;
\end{itemize}

The operator $d$ is called {\sl the exterior differential} (or
generalized gradient), the operator $\Upsilon$ is called
{\sl the Clifford differential}, and the operator $\delta$ is called
{\sl the generalized divergence}.
\bigskip

\section{ The Maxwell equations and QED equations.}
It is well known that
the Maxwell equations on Riemannian manifolds have the form
\begin{equation}
dA=F,\quad \delta F=\alpha J,
\label{Maxwell}
\end{equation}
where $A=a_\mu dx^\mu\in\la^1$, $F=\frac{1}{2}f_{\mu\nu}dx^\mu\wedge
dx^\nu\in\la^2$, $J\in\la^1$, and $\alpha$ is a constant. Using the
properties $d^2=0,\delta^2=0$ we get from (\ref{Maxwell}) that
$$
dF=0,\quad \delta J=0.
$$
If we consider the Lagrangian
$$
\L_{Maxwell}=\tr(\sqrt{-g}\,F^2)=-\frac{1}{2}\sqrt{-g}\,f_{\mu\nu}f^{\mu\nu}
$$
and take as the variational variables $a_\mu$, then we obtain
the equations (\ref{Maxwell}).
\bigskip

Now we can join the systems of equations (\ref{main}) and
(\ref{Maxwell}) and obtain the system of equations
\begin{eqnarray}
&&dx^\mu(\Upsilon_\mu\Psi+\Psi B_\mu)+A\Psi I+m\Psi
HI=0,\nonumber\\
&&\Upsilon_\mu H=[B_\mu,H],\quad \Upsilon_\mu I=[B_\mu,I],\label{QED}\\
&&H^2=1,\quad I^2=-1,\quad [H,I]=0,\nonumber\\
&&dA=F,\quad \delta F=\alpha J,\nonumber
\end{eqnarray}
where $J=\Psi\bar{\Psi}=\Psi H \Psi^*=j_\mu dx^\mu$,
$j^\mu=\tr(\bar{\Psi}dx^\mu\Psi)$, and conservative law (\ref{conserv}) can be
written with the aid of 1-form $J$ as $\delta J=0$.

In the system of equations (\ref{QED}) we
consider the differential forms $\Psi,H,I,A,F$ as unknown and the
tensor $B_\mu$ as known.
\bigskip

\pi We suppose that the system of equations (\ref{QED}) describes the
local interaction of two physical fields. Namely the field of matter
$\{\Psi,H,I\}$
(which is identified with the wave function of the electron) and the
electromagnetic field $\{A,F\}$.
So the equations (\ref{QED}) is QED equations
with presence of the gravity field $\{B_\mu,C_{\mu\nu}\}$.
\bigskip

Let us define the differential operators of the first order $\D_\mu$,
which act on tensors from $\la\top_r$
$$
\D_\mu U=\Upsilon_\mu U-[B_\mu,U]
$$
and put
$$
\D=dx^\mu\D_\mu.
$$
Using the operators $\D_\mu$ the gauge transformation
$B_\mu\to S^{-1}B_\mu-S^{-1}\Upsilon_\mu S$, $\,\,S\in\Spin$
can be written as
\begin{equation}
B_\mu\to B_\mu -S^{-1}\D_\mu S,\quad S\in\Spin.
\label{trans5}
\end{equation}
and the equations (\ref{QED}), together with the (\ref{BG}), can be
written as
\begin{eqnarray}
&&\D\Psi+A\Psi I+B\Psi+m\Psi HI=0,\nonumber\\
&&\D_\mu B_\nu-\D_\nu
B_\mu+[B_\mu,B_\nu]=\frac{1}{2}C_{\mu\nu},\nonumber\\
&&\D_\mu H=0,\quad\D_\mu I=0,\nonumber\\
&&H^2=1,\quad I^2=-1,\quad [H,I]=0,\label{UT}\\
&&dA=F,\quad\delta F=\alpha J,\nonumber
\end{eqnarray}
where $\alpha$ is a constant, $B=dx^\mu B_\mu$, and
$J=\Psi\bar{\Psi}=\Psi H\Psi^*$.

\theorem 7. The system of equations (\ref{UT}) is invariant under the
gauge transformations
\begin{eqnarray*}
&&\Psi\to\Psi S,\quad H\to S^{-1}HS,\quad I\to S^{-1}IS,\quad
B_\mu\to S^{-1}B_\mu S-S^{-1}\Upsilon_\mu S,\\
&&(A,F,C_{\mu\nu})\to(A,F,C_{\mu\nu})
\end{eqnarray*}
and
$$
\Psi\to\Psi\exp(\lambda I),\quad A\to A-d\lambda,\quad
(H,I,F,B_\mu,C_{\mu\nu})\to(H,I,F,B_\mu,C_{\mu\nu}),
$$
where $S\in\Spin$ and $\lambda\in\la^0$.

A proof follows from the theorems 3,4,6.
\bigskip

We suppose that in the system of equations (\ref{UT}) the forms
$\Psi,H,I,A,F,B_\mu$ are unknown and the forms $C_{\mu\nu}$ (the curvature
tensor) are known. As a gravity Lagrangian we may take Einstein-Hilbert
Lagrangian $\sqrt{-g}\,R$, where $R$ is the scalar curvature. Also,
there is an interesting possibility to try to describe a gravity field
using Lagrangian $\tr(\sqrt{-g}\,C_{\mu\nu}C^{\mu\nu})$. Further
development for this part of the model is needed.


\section{ Equations in Minkowski space.}
Let us suppose that Riemannian manifolds $\V$ under consideration is
such that in coordinates $x^\mu$ the metric tensor has the form
$$
\|g_{\mu\nu}\|=\|g^{\mu\nu}\|=\diag(1,-1,-1,-1).
$$
So we may consider this Riemannian manifolds as Minkowski space
admitting only Lorentzian changes of coordinates. Let $e_\mu$ be basis
coordinate vectors and $e^\mu=g^{\mu\nu}e_\nu$ be basis covectors.
Differential forms
$\frac{1}{k!}u_{\mu_1\ldots\mu_k}dx^{\mu_1}\ww dx^{\mu_k}$ can be
written as the exterior forms
$$
\frac{1}{k!}u_{\mu_1\ldots\mu_k}e^{\mu_1}\ww e^{\mu_k}.
$$
In Minkowski space ${\Gamma^\alpha}_{\mu\nu}\equiv0$ and
$$
\Upsilon_\mu V=\partial_\mu V,\quad V\in\la.
$$
The curvature tensor $R_{\mu\nu\alpha\beta}\equiv0$. The tensor
$B_\mu\in\la^2\top_1$ is such that
\begin{equation}
\partial_\mu B_\nu-\partial_\nu B_\mu-[B_\mu,B_\nu]=0.
\label{BB}
\end{equation}
The solutions of this equations are
\begin{equation}
B_\mu=U^{-1}\partial_\mu U,\quad U\in{\rm Spin(1,3)}.
\label{BM}
\end{equation}
By the Theorem 2, the covector
$$
B_\mu^\prime=S^{-1}B_\mu S-S^{-1}\partial_\mu
S=(US)^{-1}\partial_\mu(US)
$$
also satisfy (\ref{BB}).

Now we may consider the system of equations (\ref{main}) in
Minkowski space
\begin{eqnarray}
&&e^\mu(\partial_\mu\Psi+\Psi I a_\mu+\Psi B_\mu)+m\Psi
HI=0,\nonumber\\
&&\partial_\mu H=[B_\mu,H],\quad \partial_\mu I=[B_\mu,I],\label{main:Minkowski}\\
&&H^2=1,\quad I^2=-1,\quad [H,I]=0,\nonumber
\end{eqnarray}
where $\Psi\in\la^\even$, $I\in\la^2$, $H\in\la^1$, $a_\mu\in\la^0\top_1$,
and $B_\mu\in\la^2\top_1$ satisfy (\ref{BM}).
According to the Theorem 4, the system of equations
(\ref{main:Minkowski}) is
invariant under the gauge transformation (\ref{trans2}) which depends on
an exterior form $S=S(x)\in {\rm Spin(1,3)}$.

The equations (\ref{main:Minkowski}) are the new form of the tensor
Dirac equation with a ${\rm Spin(1,3)}$ gauge symmetry in Minkowski
space. A wave function of the electron is identified with the full set
$\{\Psi,H,I\}$.

If we take $S=U^{-1}$, then the gauge
transformation (\ref{trans2}) has the form
$$
\Psi\to\Psi^\prime=\Psi S,\quad
H\to H^\prime=S^{-1}HS,\quad
I\to I^\prime=S^{-1}IS,\quad
B_\mu\to 0,\quad
a_\mu\to a_\mu
$$
and the system of equations (\ref{main:Minkowski})
is identical to the tensor
Dirac equation \cite{JMP}
\begin{eqnarray}
&&e^\mu(\partial_\mu\Psi^\prime+\Psi^\prime I^\prime a_\mu)+
m\Psi^\prime H^\prime I^\prime=0,\nonumber\\
&&\partial_\mu H^\prime=0,\quad \partial_\mu I^\prime=0,\label{tde}\\
&&(H^\prime)^2=1,\quad (I^\prime)^2=-1,\quad [H^\prime,I^\prime]=0,\nonumber
\end{eqnarray}
This system of
equations is invariant under the global transformation
$$
\Psi^\prime \to\Psi^\prime V,\quad H^\prime\to V^{-1}H^\prime V,
\quad I^\prime\to V^{-1}I^\prime V,\quad a_\mu\to
a_\mu,
$$
where $V\in{\rm Spin(1,3)}$ and $\partial_\mu V=0$.
\bigskip

\section{A geometrical interpretation of the model.}
Let $\M$ be a four dimensional differentiable manifolds and let
$\V=\{\M,g_{\mu\nu}\}$ be the Rimannian manifolds with Levi-Civita
connection ${\Gamma^\lambda}_{\mu\nu}$, with the covariant derivatives
$\nabla_\mu$, with the Clifford derivatives $\Upsilon_\mu$, and with the
curvature tensor $R_{\alpha\beta\mu\nu}$ defined in previous sections.
Suppose that a new structure on $\V$ is given. Namely the affine
connection ${\check{\Gamma}^\lambda}{}_{\mu\nu}$. We get definitions of
the covariant derivatives
$\check{\nabla}_\mu$, the Clifford derivatives $\check{\Upsilon}_\mu$,
and the
curvature tensor $\check{R}_{\alpha\beta\mu\nu}$, replacing ${\Gamma^\lambda}_{\mu\nu}$
by ${\check{\Gamma}^\lambda}{}_{\mu\nu}$ in the corresponding definitions in
previous sections. We suppose that the affine connection
${\check{\Gamma}^\lambda}{}_{\mu\nu}$ is metric compatible
$$
\check{\nabla}_\kappa g_{\mu\nu}=0,\quad
\check{\nabla}_\kappa g^{\mu\nu}=0.
$$
It is convenient to introduce the tensor
$$
{K^\lambda}_{\mu\nu}={\check{\Gamma}^\lambda}{}_{\mu\nu}-{\Gamma^\lambda}_{\mu\nu},
$$
which we, following \cite{Nakahara}, will call contorsion.
It is easy to see that the affine connection is metric compatible iff
$K_{\nu\mu\lambda}=-K_{\lambda\mu\nu}$. Torsion is expressed via
contorsion as
$$
{T^\lambda}_{\mu\nu}={K^\lambda}_{\mu\nu}-{K^\lambda}_{\nu\mu}.
$$
Conversely, the contorsion of a metric compatible connection is
expressed via torsion as
$$
K^\lambda{}_{\mu\nu}=\frac{1}{2}(T^\lambda{}_{\mu\nu}+T_\mu{}^\lambda{}_\nu+
T_\nu{}^\lambda{}_\mu)
$$
(see \cite{Nakahara}, formula (7.35)).

So we arrive at the affine space
$\{\M,g_{\mu\nu},{K^\lambda}_{\mu\nu}\}$.
Let us define the tensors
\begin{eqnarray*}
b_{\alpha\beta\mu}&=&-\frac{1}{2}K_{\alpha\mu\beta},\\
B_\mu &=& \frac{1}{2}b_{\alpha\beta\mu}dx^\alpha\wedge
dx^\beta\in\Lambda^2\top_1,\\
C_{\mu\nu} &=& \frac{1}{2}R_{\alpha\beta\mu\nu}dx^\alpha\wedge
dx^\beta\in\Lambda^2\top_2.
\end{eqnarray*}

\theorem 8. If $U\in\Lambda$, then
$$
\check{\Upsilon}_\mu U=\Upsilon_\mu U - [B_\mu,U].
$$
\proof\,\, follows from the formula
$$
{K^\nu}_{\mu\lambda} dx^\lambda=[B_\mu,dx^\nu],
$$
which can be easily checked.
\bigskip

\theorem 9. (F.E.Burstall, A.D.King, N.G.Marchuk, D.G.Vassiliev)
The following equality holds
$$
\Upsilon_\mu B_\nu-\Upsilon_\nu
B_\mu-[B_\mu,B_\nu]=\frac{1}{2}C_{\mu\nu}
$$
iff
$$
\check{R}_{\alpha\beta\mu\nu}=0.
$$
\proof. Suppose that the tensors $q_{\alpha\beta\mu\nu}$ and
$\check{R}_{\alpha\beta\mu\nu}$ are such that
$$
\frac{1}{2}q_{\alpha\beta\mu\nu}dx^\alpha\wedge
dx^\beta=\Upsilon_\mu B_\nu-\Upsilon_\nu
B_\mu-[B_\mu,B_\nu]-\frac{1}{2}C_{\mu\nu}
$$
and
$$
\check{R}_{\alpha\beta\mu\nu}=g_{\kappa\alpha}(
\partial_\mu {\check{\Gamma}^\kappa}{}_{\nu\lambda}-\partial_\nu
{\check{\Gamma}^\kappa}{}_{\mu\lambda}+
{\check{\Gamma}^\kappa}{}_{\mu\eta}{\check{\Gamma}^\eta}{}_{\nu\lambda}-
{\check{\Gamma}^\kappa}{}_{\nu\eta}{\check{\Gamma}^\eta}{}_{\mu\lambda}).
$$
Then it can be easily checked that
$$
\check{R}_{\alpha\beta\mu\nu}=-2q_{\alpha\beta\mu\nu}.
$$
This completes the proof.
\bigskip

This theorem leads us to the conclusion that the equations
(\ref{UT}) can be considered as equations in the flat affine space
($\check{R}_{\alpha\beta\mu\nu}=0,\, \D_\mu=\check{\Upsilon}_\mu$).
\bigskip

\noindent {\bf Postulate}  (The flat affine field model of gravitation).
{\sl We suppose that the physical space-time is a flat affine space such
that
\begin{description}
\item[(i)] The metric tensor $g_{\mu\nu}$ satisfies conditions of the
first section.
\item[(ii)] The affine connection ${\check{\Gamma}^\lambda}{}_{\mu\nu}$ is
metric compatible and defined via contorsion ${K^\lambda}_{\mu\nu}$ or,
equivalently, by $B_{\mu}$.
\item[(iii)] The affine connection curvature
$\check{R}_{\alpha\beta\mu\nu}=0$.
\end{description}
Then the gravitation field is the pair $\{B_\mu,C_{\mu\nu}\}$, where
$B_\mu$ is identified with a potential of gravity and
$C_{\mu\nu}$ is identified with a strength of gravity.}


\end{document}